# Partially Diffusive Helium-Silica Compound in the Deep Interiors of Giant Planets

Cong Liu,[1] Junjie Wang,[1] Xin Deng,[2] Xiaomeng Wang,[1] Chris J. Pickard,[3,4] Ravit Helled,[5] Zongqing Wu,[2] Hui-Tian Wang,[1] Dingyu Xing,[1] Jian Sun[1,*]

[1] *National Laboratory of Solid State Microstructures, School of Physics and Collaborative Innovation Center of Advanced Microstructures, Nanjing University, Nanjing, 210093, P. R. China*

[2] *Laboratory of Seismology and Physics of Earth's Interior, School of Earth and Space Sciences, University of Science and Technology of China, Hefei, China*

[3] *Department of Materials Science & Metallurgy, University of Cambridge, 27 Charles Babbage Road, Cambridge CB3 0FS, UK*

[4] *Advanced Institute for Materials Research, Tohoku University 2-1-1 Katahira, Aoba, Sendai, 980-8577, Japan*

[5] *Institute for Computational Science, Center for Theoretical Astrophysics & Cosmology, University of Zurich, Switzerland*

## Abstract

Helium is the second most abundant element in the universe, and together with silica, they are major components of giant planets. Exploring the reactivity and state of helium and silica under high pressure is of fundamental importance for developing and understanding of the evolution and internal structure of giant planets. Here, using first-principles calculations and crystal structure predictions, we identify four stable phases of a helium-silica compound with seven/eight-coordinated silicon atoms at pressure range of 600-4000 GPa, corresponding to the interior condition of the outer planets in the solar system. The density of $HeSiO_2$ agrees with current structure models of the planets. This helium-silica compound exhibits a helium diffusive state at the high pressure and high temperature

---

[*] Corresponding author. jiansun@nju.edu.cn

**conditions along the isentropes of Saturn, a metallic fluid state in Jupiter, and a solid state in the deep interiors of Uranus and Neptune. The reaction of helium and silica may lead to the erosion of the rocky core of giant planets and form a diluted core region. These results highlight the reactivity of helium under high pressure to form new compounds, and also provides evidence to help build more sophisticated interior models of giant planets.**

## Introduction

Understanding the interiors of the giant planets in our Solar System is a key objective of planetary science and a multidisciplinary challenge combining condensed matter physics, astrophysics, and geophysics. This challenging task demands abundant accurate measurements, accompanied by theoretical models that are used to infer the planetary conditions and fit the available observational constraints. Traditional models[1–3] describe the outer planets (Jupiter, Saturn, Uranus, and Neptune) with three distinct layers: a gas-rich envelope mainly composed by hydrogen and helium, a denser convective and electrically conductive envelope that yields huge planetary magnetic fields (this layer is mainly composed of metallic fluid hydrogen for Jupiter and Saturn, and of superionic water/ammonia/methane mixture for Uranus and Neptune), together with a compact heavy element central core, with a density discontinuity at the core-envelope-boundary.

Thanks to updated gravity data from the Juno and Cassini missions and advance in planetary models, more complex models have been developed, in which the planetary interior is inhomogeneous and the core is diluted[4–7]. For both Jupiter and Saturn, the heavy elements could be gradually distributed or homogeneously mixed with lighter elements (hydrogen and helium) and extended to about half of the planet's total radius[6,7]. Such extended cores are difficult to explain within standard giant planet formation models[8], and therefore it is was suggested that Jupiter's fuzzy core could be a result of an energetic giant impact between a large planetary embryo with ten times Earth mass and the proto-Jupiter[9].

However, while it is still challenging to construct accurate and unique planetary structure models for the outer giant planets, we can investigate the relevant physical and chemical processes of simple elements at high pressures and high temperatures to guide planetary interiors[5,6,10]. For instance, in the gaseous planets Jupiter and Saturn, the equation of states (EOS) of hydrogen and helium in a wide range of pressures and temperatures indicated that helium is expected to demix from hydrogen, leading to helium settling toward the deep interior, known as "helium rain"[11,12]. Although helium is the most inert element at ambient pressure, in the deep interior of giant planets, the pressures and temperatures may provide sufficient energy for helium to form new compounds together with other ingredients of planet interiors. Such examples include our previously reported helium-hot ice compounds[13–15]. Among the outer giant planets in our Solar System, another major component of the heavy elements is silica, which undergoes a series of phase transitions with compression[16–18] and may have a significant influence on the formation and evolution of terrestrial planets[19]. For the outer planets, the possible mixing or even chemical reaction of helium and silica could be important.

As aforementioned, helium and silica are major components giant planets, but whether they can form new stable compounds under pressure and what states they can exist at giant planetary conditions are still open questions. To address these questions, we have systematically studied the helium silica system within the pressure range of planetary core conditions and found four new stable $HeSiO_2$ phases by a crystal structure prediction method and first-principles calculations. Further molecular dynamics simulations show that these $HeSiO_2$ compounds may survive inside Saturn with superionic-like helium diffusive phase and in Uranus and Neptune with solid phase, which provides more explication of the formation of diluted core in gaseous planets.

## Results

We explore the crystal structures in helium-silica system in the pressure range of 500-4000 GPa using variable composition structure prediction method. A structure is regarded to be thermodynamically stable if its enthalpy of formation is negative relative to the mixture of the most stable phases of solid helium (hcp) and silica (pyrite-type,

R-3, and Fe$_2$P-type)[16–18] at the corresponding pressures. As the convex hulls shown in Fig. 1(a), we find that helium and silica will form a new stable HeSiO$_2$ compound just above 600 GPa and such HeSiO$_2$ compound inclines to gain its energetic stability with further compression. We have checked the stability of the newly predicted HeSiO$_2$ compound against different exchange correlation functionals including different van der Waals (vdW) corrections and a full-potential linearized augmented plane wave (FP-LAPW) method, as shown in Figs. S1 of Supplemental Material. The results demonstrate that the HeSiO$_2$ compound can survive under deep planetary pressure.

As shown in Fig. 1 (b), we find several energetically competitive candidates for the HeSiO$_2$ compound. A *Pnma* phase (denoted as *Pnma*-I), will gain its stability at around 605 GPa, and then, it can transform into a *Pmn2$_1$* phase at ~1100 GPa. We find that the enthalpy of the *Pnma*-I and *Pmn2$_1$* phases are pretty close, because they share the same point group and have similar crystal structures, as shown in Fig. 1 (c) and (d). With increasing pressure, the *Pmn2$_1$* phase transforms into another *Pnma* phase (denoted as *Pnma*-II), at around 2100 GPa. Finally, a *Pnma*-III phase, would gain its stability above 2300 GPa. Phonon calculations indicate that the above structures are all dynamically stable, as shown in Fig. S2 of Supplemental Material. We also calculate the electronic band gaps of these HeSiO$_2$ compounds and they are completely insulating with wide band gaps between that of He and SiO$_2$ crystal, as shown in Fig. S3 of Supplemental Material. It is clear that the electronic band gaps incline to decrease in He and SiO$_2$ crystals under increasing compression. The SiO$_2$ crystal is predicted to transform into semi-conductor at 4 TPa and could metallize at higher pressure. Whereas the HeSiO$_2$ compounds have an abnormal pressure dependence with an increase in the band gap over the pressure range 500–2000 GPa. Previous work[18] showed that six-fold silica would transform into a mixed coordination silica with an averaged coordination number of 8 by compression. Interestingly, in our newly found HeSiO$_2$ compound, all silicon atoms in *Pnma*-I, *Pmn2$_1$*, and *Pnma*-II phase are seven-fold coordinated with oxygen atoms, while *Pnma*-III phase is purely eight-fold. These seven-fold and eight-fold configurations are rare in pure silica, which indicates that the inserting of helium atoms can significantly changes the packing of silicon and oxygen atoms, as well as the silicon-oxygen bonding behaviors. Meanwhile, we also find that a purely eight-fold HeSiO$_4$ compound with *I422* phase can survive above 400 GPa, as shown in Fig. S4 of Supplemental Material. This may shed light on further explorations on chemical coordination in helium compounds.

Since HeSiO$_2$ compounds could exist in high pressure range corresponding to deep interiors of giant planets, we investigated the equation of state of these HeSiO$_2$ compounds, as shown in Fig. 2. In the traditional 3-layer models for of Jupiter and Saturn with distinct layers, their density curves increase sharply at core mantle boundary. Our *ab initio* calculations show that the density curves of the HeSiO$_2$ compounds have a smooth tendency: increasing by compression, while decreasing by heating. Most importantly, they are located just between that of the core and mantle, which indicates that these newly found HeSiO$_2$ compounds very possibly exist near the core mantle boundary of giant gaseous planets, such as Jupiter and Saturn. Especially when consider diluted core models, helium/hydrogen is expected to erode the core and their density curves would change to a smooth one, which should agree with the density of states of these HeSiO$_2$ compounds even more closely, in the deep interiors of Jupiter and Saturn.

In order to gain a better understanding of such HeSiO$_2$ compounds affecting the interior model of giant planets, one should account for both the equation of states calculations and the pressure-temperature phase diagrams, because the internal structure models must be consistent with the phase diagram of the assumed materials and their dynamical behavior. We perform extensive AIMD simulations at deep planetary conditions to study the dynamical properties of our predicted HeSiO$_2$ compounds. Diffusion coefficients (*D*) were calculated for the silicon, oxygen, and helium atoms from the averaged mean-square displacements (MSD) to monitor different types of atomic motions ($D = \partial \text{MSD}/\partial t$). We demonstrated the atomic trajectories from simulations at the initial pressure of 600 GPa for the *Pnma*-I phase HeSiO$_2$ compound as an example, as shown in Fig. S5 of Supplemental Material. At 5000 K, all atoms oscillate around their equilibrium positions, resulting in three horizontal MSD curves with slightly oscillating ($D = 0$). When heating up to 9000 K, we find that the helium atoms became diffusive freely within the static silica frameworks during the simulations ($D_{He} > D_{Si} = D_O = 0$). These are clearly two different states: the solid phase at 5000 K and the superionic-like partial diffusive phase at 9000 K.

Inspired by this helium diffusive state in the $HeSiO_2$ compounds, we extended the pressure and temperature range of our AIMD simulations to explore the states of these $HeSiO_2$ compounds at the deep interior condition of Saturn and Jupiter, as shown in Fig. 3. Each colored symbol corresponds to an independent simulation to avoid correlation effects, and the pressures and temperatures were obtained from simulations by statistically averaging. Due to the different stable pressure range of $HeSiO_2$ phases, as well as the core mantle boundary conditions of Saturn (1 TPa and 10,000 K) and Jupiter (4 TPa and 20,000 K) varying a lot, we separately simulate $HeSiO_2$ compounds in pressure range of 500-1200 GPa with *Pnma*-I phase (as shown in Fig. 3 (a)) and in pressure range of 2000-5000 GPa with *Pnma*-III phase (as shown in Fig. 3 (b)) up to the melting temperature. It is clear that the superionic-like helium diffusive state is widespread at high pressure and high temperature and exists between the solid phase and fluid phase in the $HeSiO_2$ phase diagram. With increasing pressure, this helium diffusive state appears at higher temperatures.

For comparison, the isentropes of the giant outer planets (Jupiter, Saturn, Uranus, and Neptune) are also plotted in Fig. 3 to represent the pressure temperature profiles at deep interior conditions. We find that the pressure temperature profile of Saturn (magenta) perfectly passes through the superionic-like helium diffusive region of the $HeSiO_2$ compound, which suggests that such helium diffusive $HeSiO_2$ compounds can exist near the core mantle boundary of Saturn. As for Jupiter, both adiabatic (red) and non- adiabatic (white) pressure temperature profiles are slightly higher than the helium diffusive region. In the other words, the $HeSiO_2$ compounds are totally melted in the core region, if we account for a diluted core model of Jupiter. Previous works[20,21] on typical mantle silicates (MgO, $SiO_2$, and $MgSiO_3$) showed that upon melting, the behavior of $SiO_2$ changes from semi-conducting to semi-metallic, indicating a magnetic dynamo process would develop in the magma oceans of Super-Earths. We calculate the electronic band gap of HeSiO$_2$ compounds at different temperatures, as shown in Fig. S6 of Supplemental Material. It is clear that the band gap decreases with increasing temperature. When it enters the helium diffusive region in the diluted core region of Saturn, the band gap sharply decreases. At higher temperature HeSiO$_2$ compounds may transform into metallic with zero-bandgap in fluid state, which may affect the conductivity and therefore the magnetic field generation in Jupiter. The pressure temperature profiles of Uranus and Neptune are much lower than that of Jupiter and Saturn, and in this case the HeSiO$_2$ compounds may exist in solid form in the deep interior of Uranus and Neptune when helium deposits and erode the core.

We also calculate the elastic and wave velocity properties of the HeSiO$_2$ compound to explore how it affects the evolution of the interior of Uranus and Neptune. The calculated bulk moduli ($K_S$), shear moduli (G), compressional velocities ($V_P$), and shear wave velocities ($V_S$) of the *Pnma* phase HeSiO$_2$ compound, as well as pure He and silica for comparison, at pressures range of 500-1000 GPa are shown in Fig. 4. He and HeSiO$_2$ compound both have almost linear temperature and pressure dependences of bulk moduli and shear moduli. For instance, in the HeSiO$_2$ compound, the first temperature and pressure derivate of elastic properties almost keep unchanged ($\partial K_S/\partial T$ = -0.013 GPa/K, $\partial G/\partial T$ = -0.016 GPa/K, $\partial K_S/\partial P$ = 2.7, and $\partial G/\partial P$ = 1.1). In contrast, noticeable nonlinear dependences on temperature and pressure are observed for the wave velocities. Due to the phase transitions occurring in the silica pressure-temperature phase diagram, there are several discontinuities in elastic curves of silica.

The chemical composition can also affect the elastic moduli and wave velocities of these planetary matters. Although pure He has a very small bulk moduli and shear moduli, He has much larger compressional velocities and shear wave velocities than silica and HeSiO$_2$ due to its small density ($V_P = \sqrt{(K_S + \frac{4}{3}G)/\rho}$ and $V_S = \sqrt{G/\rho}$, where $\rho$ is density). In silica, the Cotunnite-type phase surviving at high temperature region has a small shear moduli and shear wave velocities with respect to other phases, which indicates that an ultralow-velocity zones may exist in the mantle of super-earth. Compared to silica, HeSiO$_2$ has a slightly smaller bulk moduli but larger shear moduli.

Due to the decrease of density by inserting helium, the velocities of HeSiO$_2$ compound have a larger compressional velocities and shear wave velocities than that of silica. Notably, the temperature does not significantly affect wave velocities. For instance, wave velocities vary 0.16 and 0.23 km/s for compressional velocities and shear wave velocities from 4000 K to 7000 K at 500 GPa and these differences are evidently diminished at high pressure, which decrease to 0.08 and 0.14 km/s at 1000 GPa. Thus, the inserting of helium may evidently increase the wave velocities of silica and our newly predicted HeSiO$_2$ may affect the model of the deep interior of Uranus and Neptune.

## Discussion

The composition of giant planets cannot be measured directly. Instead, their bulk compositions and internal structures must be inferred indirectly from interior models that fit the available measured physical parameters. Here we used calculated equations of state (EoS) of our predicted HeSiO$_2$ compounds to derive the density (as shown in Fig. 2.) and the associated pressure temperature phase diagram (as shown in Fig. 3) and thus we can sketch internal structure models for Jupiter, Saturn, Uranus, and Neptune, as shown in Fig. 5. The gas giants, Jupiter and Saturn, are mainly composed by hydrogen and helium, while Uranus and Neptune are expected to consist of large fractions of water, ammonia, and methane, although their exact compositions are far from being well-constrained[5].

Our static calculations show that helium and silica can react with each other in the deep interiors of giant planets, and the HeSiO$_2$ compounds may survive near the core mantle boundary of giant planets. Especially considering that helium may be immiscible with hydrogen at 1-2 MBar, namely "helium rain", it is expected to settle down into deep interior of Jupiter and Saturn. Thus, helium in the deep interior could erode the compact heavy-element core, resulting in a gradually expanding core region and forming a diluted one, as shown in Fig. 5 (a) and (b). Pressure temperature phase diagram indicates that the HeSiO$_2$ compounds may exist at the diluted core region of Saturn in helium diffusive state. According to the planetary models[6], our predicted helium diffusive HeSiO$_2$ compounds are buried beneath 0.33 R in Saturn, while HeSiO$_2$ compounds may transform into metallic fluid in Jupiter. For Uranus and Neptune, our previous work[13–15] suggested that hot ice layer could mix with helium and cause helium settling into the deep interior. If the Uranus and Neptune indeed have composition gradients, helium (and hydrogen) could exist also in the planetary deep interiors. If helium erodes the silica core, HeSiO2 compounds may even exist at the diluted core region of Uranus and Neptune in solid state, corresponding to 0.15 R and 0.31 R, respectively. The elastic calculations also suggests that helium may increase the wave velocity in the core region of Uranus and Neptune with the compressional wave velocities of 20.3 km/s and shear wave velocities of 10.4 km/s at 600 GPa. When going deep into inner core, the wave velocities will increase to 21.9 km/s for compressional velocities and 11.1 km/s for shear wave velocities at 800 GPa. Here, we only fitted density curves and pressure temperature phase diagram of these HeSiO$_2$ compounds to the representative planetary models, more physical data (such as their masses, radii, gravitational and magnetic fields, 1-bar temperatures, atmospheric composition, and internal rotations) of planets are required to construct a more comprehensive model. Moreover, it should be noted that the hydrogen can also affect the stability of silica at high pressure, as we discussed in another paper[22]. However, a complete investigation in the H-He-Si-O system is beyond the scope of this work.

In summary, using crystal structure prediction and *ab initio* calculations, we have predicted four phases of HeSiO$_2$ compound (*Pnma*-I, *Pmn2$_1$*, *Pnma*-II, and *Pnma*-III) which can gain their stability at pressure range of 600-4000 GPa, corresponding to deep interior conditions of giant planets, such as Jupiter, Saturn, Uranus, and Neptune. Due to the spatial occupation of helium atoms, the silicon atoms in these compounds are all bonded to seven/eight oxygen atoms, which are very rare cases in pure silica. Thus, helium and other inert gas atoms can be used as a space filler to design compounds with usual chemical bonding and coordination under high pressure. Equation of states calculations suggest that the density curves of our newly predicated HeSiO$_2$ compound are close to current models of Jupiter and Saturn, especially when accounting for diluted core models. Furthermore, extensive *ab initio* molecular dynamics simulations illustrated that the HeSiO$_2$ compound can survive in the helium diffusive state at the core-mantle boundary conditions of Saturn and metallic fluid state at the core-mantle boundary conditions of Jupiter, which may shed light on the formation or evolution of diluted core of such gaseous giants. While in the pressure-temperature conditions in the deep interiors of Uranus and Neptune, the HeSiO$_2$ compound is found to be in solid form due to lower temperatures. We also carried out elastic and wave velocity properties calculations for the HeSiO$_2$ compound in the pressure range of 500-1000 GPa and found that the inserting of helium increases the compressional and shear wave velocities in the core region of Uranus and Neptune. Our findings can be used to guide giant planet interior models and to significantly improve our understanding on giant planets in our solar system and beyond.

## Methods

We used Magus (machine learning and graph theory assisted universal structure searcher) code to search for the crystal structures, in which we employed the Bayesian optimization[23] and graph theory[24] to improve the search efficiency and diversity. We performed extensive crystal structure searches on He$_x$(SiO$_2$)$_y$ (x=1–4, y=1-4) at 500, 1000, 2000, and 4000 GPa with maximum atom number up to 40. Some compositions, are further double checked with extensive fixed composition searches. Each search runs over 25 generations and each generation has a population size of 60 structures. 40%-60% of the parents for the evolution of next generation are from the lowest enthalpy structures of the last generation and the left seeds are randomly produced. We also cross checked the searching results with AIRSS[25,26] combined with CASTEP[27].

DFT calculations were performed with the Vienna ab initio simulation package (VASP)[28], accompanied with the projector augmented-wave (PAW) method[29]. We chose 3s$^2$3p$^2$, 2s$^2$2p$^4$, and 2s$^2$ as valence electrons for Si, O, and He, and used the generalized gradient approximation (GGA) in the Perdew-Burke-Ernzerhof (PBE) exchange correlation functional[30]. All predicted structures were further optimized by the hard version pseudopotentials and employed a plane wave cutoff of 1050 eV and dense Monkhorst-Pack k-point sampling grids with resolutions of $2\pi \times 0.025$ leading to ionic and cell optimizations with energy and force convergences better than $10^{-6}$ eV and 0.002 eV/Å, respectively.

Elastic properties at high pressure and temperature are calculated by cij package[31] based on phonon spectrum from the PHONONPY package[32]. The static elastic constants are calculated by stress strain method.

We adopted the canonical *NVT* ensemble using a Nose-Hoover thermostat[33] to perform *ab initio* molecular dynamics (AIMD) simulations in a supercell with 192 atoms for both *Pnma*-I phase and *Pnma*-III phase, with Γ-centered k-points sampling, a normal version pseudopotentials and a cutoff energy of 720 eV were adopted to ensure energy convergence of better than $10^{-5}$ eV. Each simulation lasts for 12 ps with a time step of 1 fs, and we allowed the first 2 ps for thermalization.

## Acknowledgements

J.S. gratefully acknowledges the financial support from the National Natural Science Foundation of China (grant nos. 11974162 and 11834006), and the Fundamental Research Funds for the Central Universities. The calculations were carried out using supercomputers at the High Performance Computing Center of Collaborative Innovation Center of Advanced Microstructures, the high-performance supercomputing center of Nanjing University.

## Competing interests

CJP is an author of the CASTEP code, and receives royalty payments from its commercial sales by Dassault Systemes.

## Additional information

Correspondence should be addressed to J. S. (Email: jiansun@nju.edu.cn)

Supplementary information is available for this paper at https://xxx.xxx-xxx

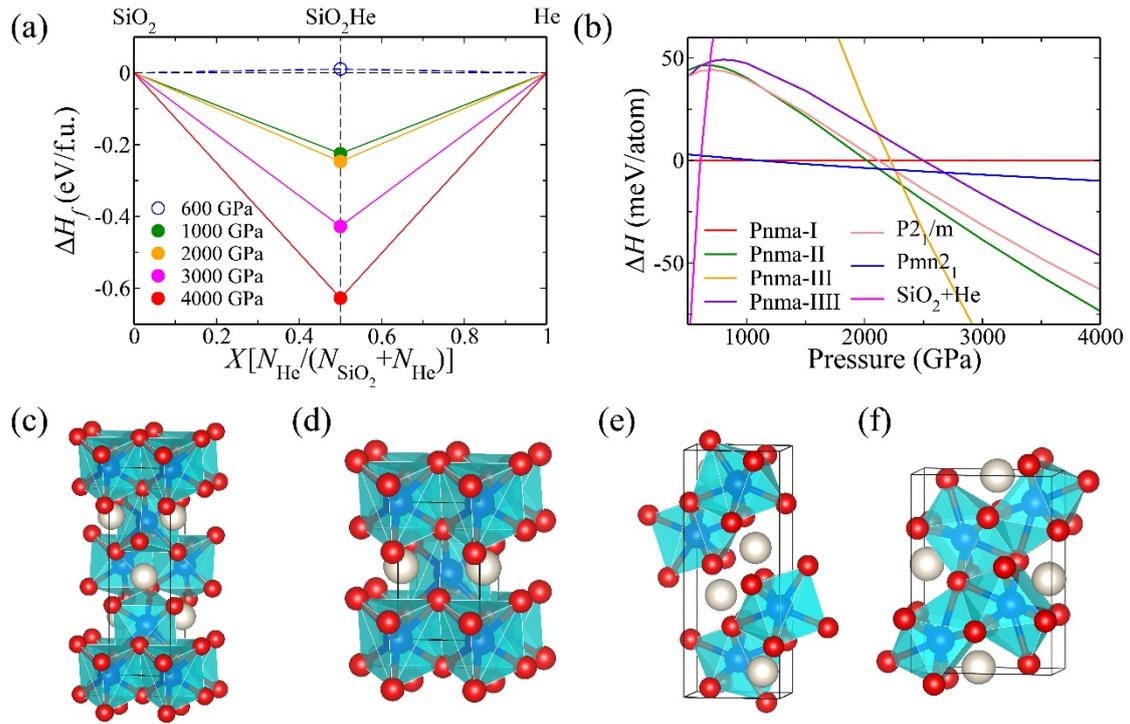

**Fig. 1. Energetics of the He-SiO₂ system and crystal structures of the most stable compounds.** (a) Convex hulls for formation enthalpies ($\Delta H_f$, relative to the most stable phases of He and SiO$_2$ [16–18]) at different pressures. (b) Enthalpies of the HeSiO$_2$ compounds of interest, as well as He-silica mixture, relative to the *Pnma*-I phase in the pressure range between 500 and 4000 GPa. (c-f) Crystal structure of the predicted stable HeSiO$_2$ compounds: (c) *Pnma*-I, (d) *Pmn2₁*, (e) *Pnma*-II and (f) *Pnma*-III, where blue, red, and silver spheres represent Si, O and He atoms, respectively.

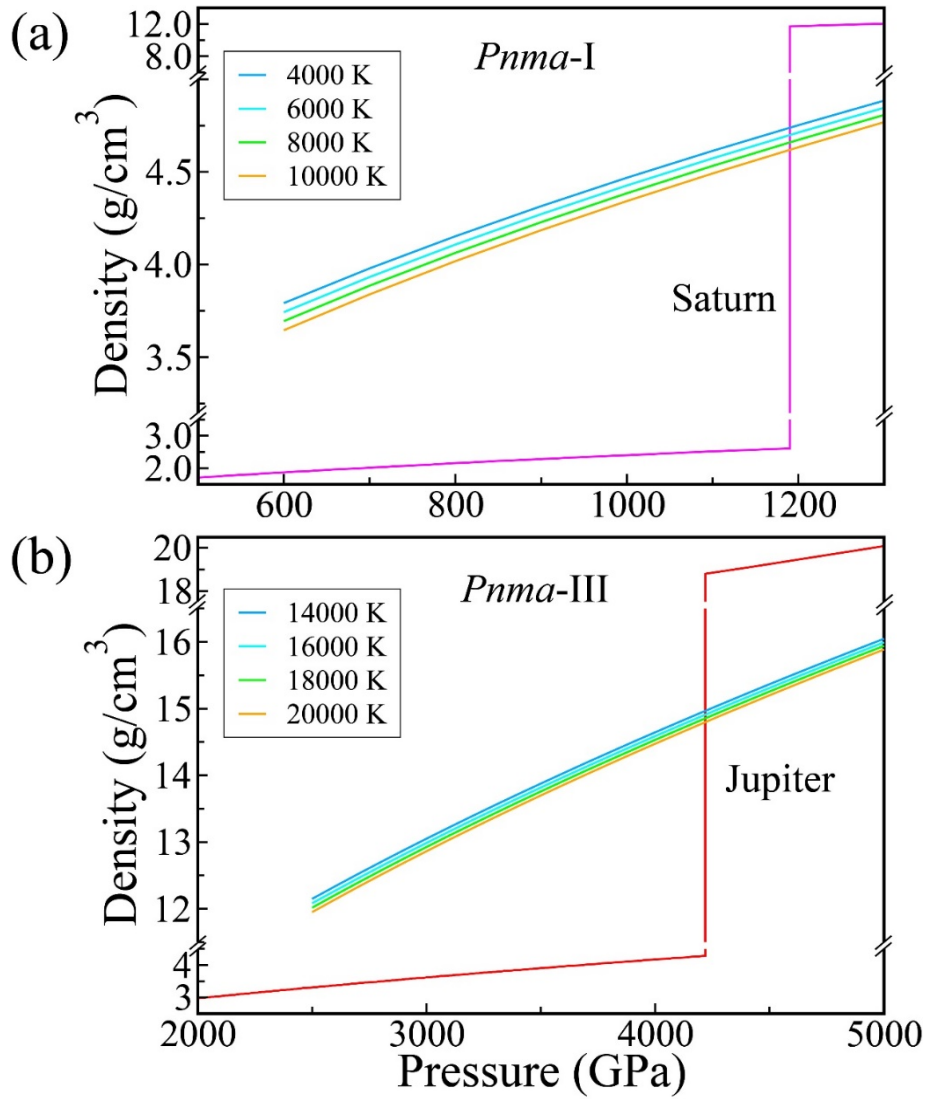

**Fig. 2. Compression curves of the He-SiO₂ compounds.** (a) *Pnma*-I phase and (b) *Pnma*-III phase. Colorful continuous lines represent *ab initio* results at variable temperatures, while the discontinuous lines represent density-pressure curves of different planetary interiors: Jupiter[34] (red), Saturn[35] (magenta).

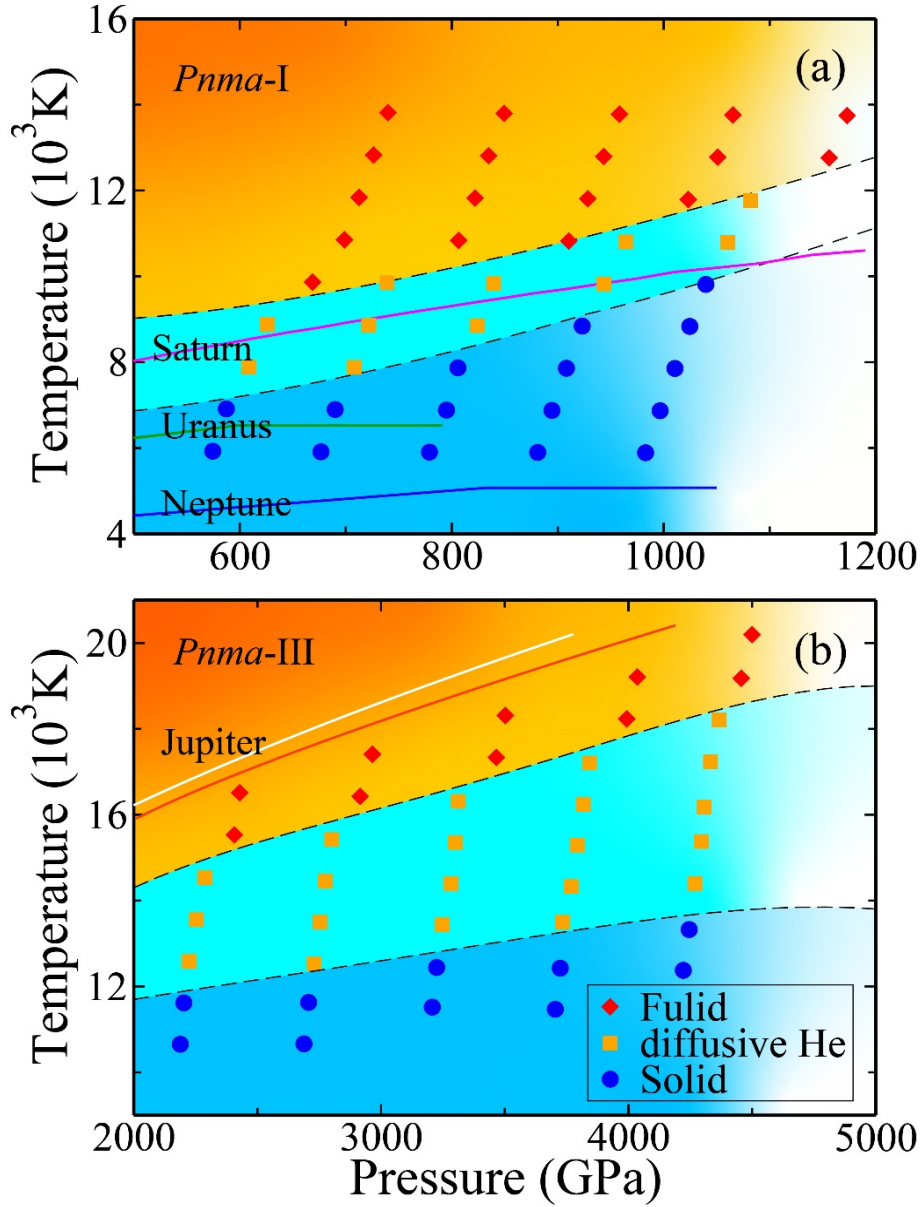

**Fig. 3. Proposed phase diagram of the HeSiO$_2$ compound**: (a) *Pnma*-I phase and (b) *Pnma*-III phase at planetary interior conditions by AIMD simulations. The simulations are marked with three different symbols: blue circle, orange square, and red diamond represent the solid, helium diffusive, and fluid states, respectively. Black dashed lines are fitted to the phase transition boundaries. The pressure-temperature profiles for giant planets are plotted in red (Jupiter[34]), magenta (Saturn[35]), dark green (Uranus[36]), and blue (Neptune[36]) for reference, assuming adiabatic interiors. A profile for the non-adiabatic Jupiter model (white)[4] is also provide for comparison.

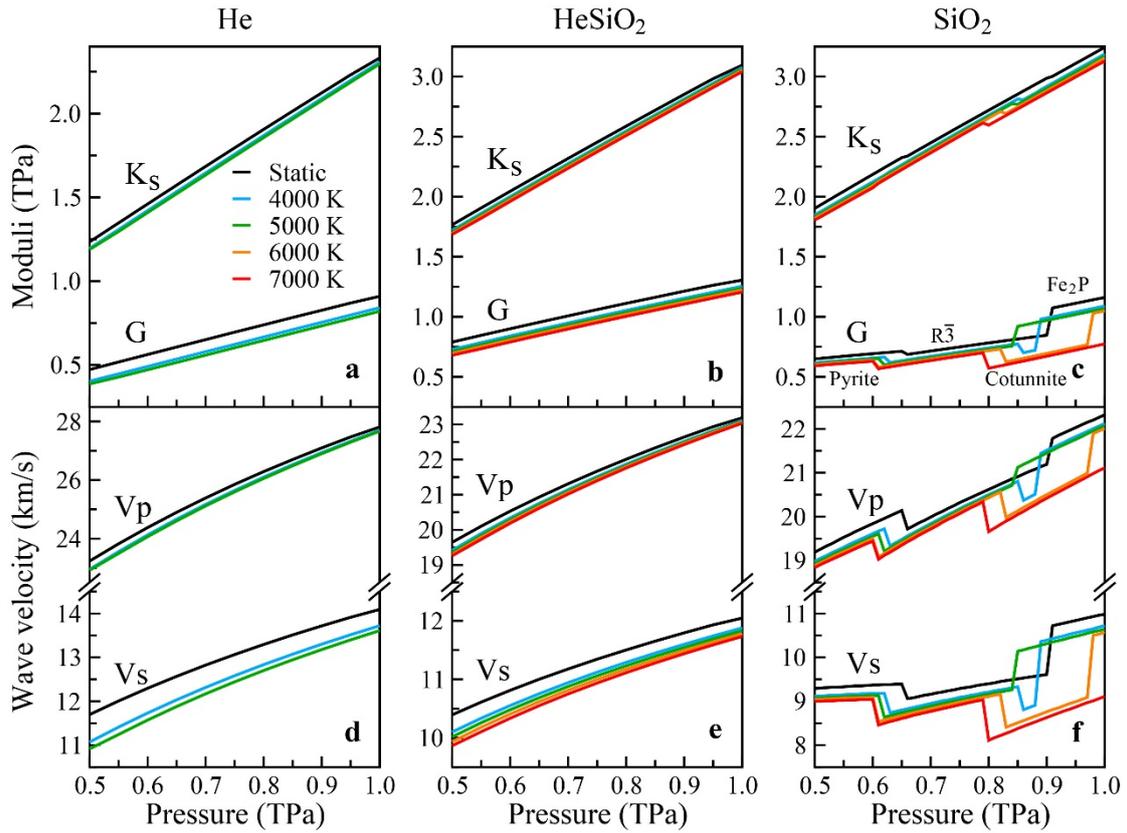

**Fig. 4. Comparisons of elastic moduli and wave velocities** between pure He, the *Pnma* phase HeSiO$_2$ compound, and SiO$_2$ along the pressure-temperature profiles for Uranus and Neptune. Elastic moduli and wave velocities. (a–c) bulk and shear moduli ($K_S$ and G), (d–f) compressional and shear wave velocities ($V_P$ and $V_S$). Colorful lines represent *ab initio* results at variable temperatures.

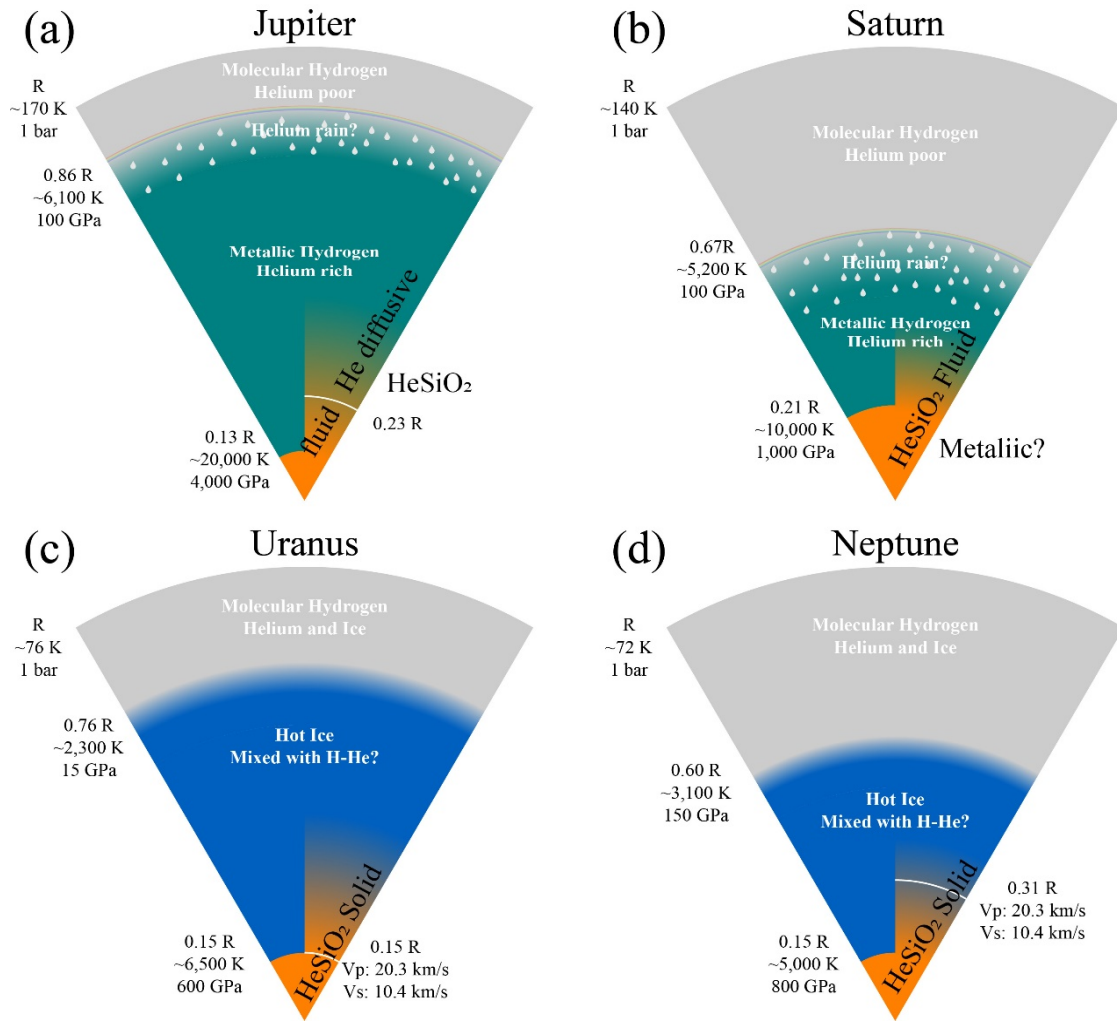

**Fig. 5**. **Sketches of the internal structures of giant planets**: (a) Jupiter, (b) Saturn, (c) Uranus, and (d) Neptune. For each planet, two possible models are shown: compact core (left) and diluted core (right). The compact core model has well-defined layers and distinct cores and the typical pressures and temperatures are denoted alongside[34–36], while the diluted core model have composition gradients and cores that are less well defined.